\documentclass[pre,showpacs,floatfix,twocolumn]{revtex4-1}
\usepackage{amssymb}
\usepackage{graphicx}
\usepackage{epsfig}

\begin{document}
\title{Generalized Metropolis dynamics with a generalized master equation: An approach for time-independent and time-dependent Monte Carlo simulations of generalized spin systems}
\author{\firstname{Roberto} \surname{da Silva}}
\email{rdasilva@if.ufgrs.br}
\affiliation{Instituto de Fisica, Universidade Federal do Rio Grande do Sul,
Av. Bento Gon\c{c}alves, 9500 - CEP 91501-970, Porto Alegre, Rio Grande do
Sul, Brazil}

\author{\firstname{Jos\'e} Roberto \surname{Drugowich de Fel\'{\i}cio}}
\email{drugo@usp.br}
\affiliation{ Faculdade de Filosofia, Ci\^{e}ncias e Letras de Ribeir\~{a}o
Preto, Universidade de S\~{a}o Paulo, Avenida Bandeirantes, 3900 - CEP
14040-901, Ribeir\~{a}o Preto, S\~{a}o Paulo, Brazil}

\author{\firstname{Alexandre} Souto \surname{Martinez}}
\email{asmartinez@ffclrp.usp.br}
\affiliation{ Faculdade de Filosofia, Ci\^{e}ncias e Letras de Ribeir\~{a}o
Preto, Universidade de S\~{a}o Paulo, Avenida Bandeirantes, 3900 - CEP
14040-901, Ribeir\~{a}o Preto, S\~{a}o Paulo, Brazil}%
\affiliation{
Instituto Nacional de Ci\^encia e Tecnologia em Sistemas Complexos, Brazil}

\date{\today}

\begin{abstract}
The extension of Boltzmann-Gibbs thermostatistics, proposed by Tsallis,
introduces an additional parameter $q$ to the inverse temperature $\beta$.
Here, we show that  a previously introduced generalized Metropolis
dynamics to evolve spin models is not local and does not obey the detailed
energy balance. In this dynamics, locality is only retrieved for $q=1$,
which corresponds to the standard Metropolis algorithm. Non-locality implies
in very time consuming computer calculations, since the energy of the whole
system must be reevaluated, when a single spin is flipped. To circumvent
this costly calculation, we propose a generalized master equation, which
gives rise to a local generalized Metropolis dynamics  that obeys the
detailed energy balance. To compare the different critical values obtained
with other generalized dynamics, we perform Monte Carlo simulations
in equilibrium for Ising model. By using the short time
non-equilibrium numerical simulations, we also calculate for this model: the
critical temperature, the static and dynamical critical exponents as
function of $q$. Even for $q\neq 1$, we show that suitable time evolving
power laws can be found for each initial condition. Our numerical experiments 
corroborate the literature results, when we use  non-local dynamics, showing that
short time parameter determination  works also in this case. However, the
dynamics governed by the new master equation leads to  different results
for critical temperatures and also  the critical exponents affecting
universality classes. We further propose a  simple algorithm to optimize
modeling the time evolution with a power law considering in 
 a log-log plot  two successive refinements.  
\end{abstract}

\pacs{05.10.Ln, 05.70.Ln, 02.70.Uu}
\maketitle

\section{Introduction}

The study of the critical properties of  magnetic systems  plays an important
role in statistical mechanics and as a consequence also  in thermodynamics. For
equilibrium, the extensitivity of the entropy is a question of principle for
 most physicists. Nevertheless, an important issue may be raised.
While many physicists believe that  statistical mechanics generalizations
 with an extra parameter $q$~\cite{tsallis_1} are suitable to study the
 optimization combinatorial process as for example the
simulated annealing (see e.g. \cite{Stariolo},\cite{Uli1997}), or areas such  as 
econophysics~\cite{martinez1,martinez2}, population
dynamics and growth models~\cite{martinez3,martinez4,martinez5,martinez6},
Bibliometry~\cite{martinez7} and others.

In this paper, we generate the critical dynamics of Ising systems
using a new master equation. This master equation leads to a generalized
Metropolis prescription, which depends only on the  spin interaction
energy variations with respect to its neighborhood. Furthermore, it
satisfies the detailed energy balance condition and it converges
asymptotically to the generalized Boltzmann-Gibbs weights. In
Refs.~\cite{Crokidakis2009,Boer2011} generalized prescriptions have been
treated as local. Here, we demonstrate that they are instead  non-local.
However,  a non-local prescription  such as
the one of Ref.~\cite{Penna1999} is numerically more expensive and  destroys the phase
transition. Another possibility is to recover locality. Using a special
deformation of the master equation, we show how to recover locality for a
generalized prescription and additionally recovering the detailed energy
balance in equilibrium spin systems, maintaining the system phase
transition. 

To apply our Metropolis prescription, we have simulated a two dimensional
Ising system in two different ways: using equilibrium Monte Carlo (MC)
simulations we estimate critical temperatures for different $q$-values and
performing time-dependent simulations. In the second part, we also
calculate the critical exponents set corresponding to each 
critical temperature.Finally,   we have 
developed an alternative methodology to refine the determination of the
critical temperature. Our approach is based on the optimization of the
magnetization power laws in log scale via of maximization of determination
coefficient ($r$) of the linear fits.

Our presentation is organized as follows. In Sec.~\ref{sec:critical_dynamics}%
, we briefly review the results of the critical dynamics for spins systems.
In this  review, we calculate the critical exponents for the several
spin phases, that emerge from different initial conditions. In Sec.~\ref%
{sec:generalized_master_equation}, we propose a new master equation that
leads to a Metropolis algorithm, which preserves locality and detailed
energy balance, also for $q\neq 1$. In Sec.~\ref{sec:results}, we simulate
an equilibrium Ising spin system in a square lattice and show the
differences between the results of our approach and of Refs.~\cite%
{Crokidakis2009,Boer2011}. Next, we evolve a Ising spin system in a square
lattice, from ordered and disordered initial conditions in the context of
time dependent simulations. From such non equilibrium Monte Carlo
simulations, also called short time simulations, we are able to calculate
the dynamic and static critical exponents ones. Finally, the conclusions are
presented in Sec.~\ref{sec:conclusions}.

\section{Critical dynamics of spin systems and time dependent simulations}

\label{sec:critical_dynamics}

Here, we briefly review finite size scaling in the dynamics relaxation of
spin systems. We present our alternative deduction of the some expected
power laws in the short time dynamics context. Readers, which want a more
complete review about this topic, may want to read  \cite{Zheng1998}. 

This topic is based on time dependent simulations, and it constitutes an
important issue in the context of phase transitions and critical phenomena.
Such methods can be applied not only to estimate the critical parameters in
spin systems, but also to calculate the critical exponents (static and
dynamic ones) through  different scaling relations by setting different
initial conditions.

The study of the statistical systems dynamical critical properties has
become simpler in nonequilibrium physics after the seminal ideas of Janssen,
Schaub and Schmittmann~\cite{Janssen1989} and Huse~\cite{Huse1989}.
quenching systems from high temperatures to the critical one, they have shown 
universality and scaling behavior to appear already in  the early stages of time evolution, 
via renormalization group techniques and numerical calculations respectively. Hence, using 
short time dynamics, one can often  circumvent the well-known problem of the critical 
slowing down that plagues investigations of  the long-time regime.

The dynamic scaling relation obtained by Janssen \textit{et al.} for the
magnetization \textit{k}-th moment, extended to finite size systems, is
written as 
\begin{equation}
\langle M^{k}\rangle (t,\tau ,L,m_{0})=b^{-k\beta /\nu }\langle M^{k}\rangle
(b^{-z}t,b^{1/\nu }\tau ,b^{-1}L,b^{x_{0}}m_{0})\text{,}
\label{mainshorttime}
\end{equation}%
where the arguments are: the time $t$; the reduced temperature $\tau
=(T-T_{c})/T_{c}$, with $T_{c}$ being the critical one, the lattice linear
size $L$ and initial magnetization $m_{0}$. Here, the operator $\langle
\ldots \rangle $ denotes averages over different configurations due to
different possible time evolution from each initial configuration compatible
with a given $m_{0}$. On the equation right-hand-side, one has: an arbitrary
spatial rescaling factor $b$; an anomalous dimension $x_{0}$ related to $%
m_{0}$. The exponents $\beta $ and $\nu $ are the equilibrium critical
exponents associated with the order parameter and the correlation length,
respectively. The exponent $z$ is the dynamic one, which characterizes the
time correlations in equilibrium. After the scaling $b^{-1}L=1$ and at the
critical temperature $T=$ $T_{c}$, the first ($k=1$) magnetization moment
is: $\langle M\rangle (t,L,m_{0})=L^{-\beta /\nu }\langle M\rangle
(L^{-z}t,L^{x_{0}}m_{0})$.

Denoting $u=tL^{-z}$ and $w=L^{x_{0}}m_{0}$, one has: $\langle M\rangle
(u,w)=\langle M\rangle (L^{-z}t,L^{x_{0}}m_{0})$. The derivative with
respect to $L$ is: $\partial _{L}\langle M\rangle =(-\beta /\nu )L^{-\beta
/\nu -1}\langle M\rangle (u,w)+L^{-\beta /\nu }[\partial _{u}\langle
M\rangle \partial _{L}u+\partial _{w}\langle M\rangle \partial _{L}w]$,
where we have explicitly: $\partial _{L}u=-ztL^{-z-1}$ and $\partial
_{L}w=x_{0}m_{0}L^{x_{0}-1}$. In the limit $L\rightarrow \infty $, $\partial
_{L}\langle M\rangle \rightarrow 0$, one has: $x_{0}w\partial _{w}\langle
M\rangle -zu\partial _{u}\langle M\rangle -\beta /\nu \langle M\rangle =0$.
The separability of the variables $u$ and $w$ in $\langle M\rangle
(u,w)=M_{1}(u)M_{2}(w)$ leads to $x_{0}wM_{2}^{\prime }/M_{2}=\beta /\nu
+zuM_{1}^{\prime }/M_{2}$, where the prime means the derivative with respect
to the argument. Since this equation left-hand-side depends only on $w$ and
the right-hand-side depends only on $u$, they must be equal to  a constant $c$.
Thus, $M_{1}(u)=u^{(c/z)-\beta /(\nu z)}$ and $M_{2}(w)=w^{c/x_{0}}$,
resulting in $\left\langle M\right\rangle (u,w)=m_{0}^{c/x_{0}}L^{\beta /\nu
}t^{(c-\beta /\nu )/z}$. Returning to the original variables, one has: $%
\langle M\rangle (t,L,m_{0})=m_{0}^{c/x_{0}}t^{(c-\beta /\nu )/z}$.

On one hand, choosing $c=x_{0}$ and calculating $\theta =(x_{0}-\beta /\nu )/z$,
at criticality ($\tau =0$), we obtain $\langle M\rangle _{m_{0}}\sim
m_{0}t^{\theta }$  corresponding to a regime under small initial
magnetization. This can be observed by a finite time scaling $%
b=t^{1/z}$ in equation \ref{mainshorttime}, at critical temperature ($\tau
=0 $) which leads to $\left\langle M\right\rangle (t,m_{0})=t^{-\beta /(\nu
z)}\langle M\rangle (1,t^{x_{0}/z}m_{0})$. Defining  $x=t^{x_{0}/z}m_{0}$, an
expansion of the averaged magnetization around $x=0$ results in: $\langle
M\rangle (1,x)=\langle M\rangle (1,0)+\left. \partial _{x}\langle M\rangle
\right\vert _{x=0}x+\mathcal{O}(x^{2})$. By construction $\langle M\rangle
(1,0)=0$, since $u=t^{x_{0}/z}m_{0}\ll 1$ and $\left. \partial _{x}\langle
M\rangle \right\vert _{x=0}$ is a constant. So, by discarding the quadratic
terms we obtain the expected power law behavior $\langle M\rangle
_{m_{0}}\sim m_{0}t^{\theta }$. This anomalous behavior of initial
magnetization is valid only for a characteristic time scale $t_{\max }$ $%
\sim m_{0}^{-z/x_{0}}$.

On the other hand, the choice $c=0$ corresponds to a case where the system
does not depend on the  initial trace of the system; and  $m_{0}=1$ leads to simple
power law: 
\begin{equation}
\langle M \rangle _{m_{0}=1}\sim t^{-\beta/(\nu z)}  \label{decay_ferro}
\end{equation}
that similarly corresponds to decay of magnetization for $t>t_{\max }$ of a
system that previously evolved from a initial small magnetization $(m_{0})$,
and had its magnetization increased up to a magnetization peak.

For  $m_{0}=0$, it is not difficult to show that the magnetization
second moment is 
\begin{equation}
\left\langle M^{2}\right\rangle _{m_{0}=0}\sim t^{(d-2\beta /\nu )/z}\;,
\end{equation}%
where $d$ is the system dimension.

Using Monte Carlo simulations,  many authors have obtained the dynamic
exponents $\theta $ and $z$ as well as the static ones $\beta $ and $\nu $,
and other specific exponents for many different models and situations:
Baxter-Wu~\cite{Arashiro2003}, 2, 3 and 4-state Potts~\cite%
{daSilva2002a,daSilva2004}, Ising with multispin interactions~\cite%
{Simoes2001}, models with no defined Hamiltonian (celular automata and
contact process)~\cite{daSilva2004a,tome1998,daSilva2005}, models with
tricritical point~\cite{daSilva2002}, Heisenberg~\cite{Fernandes2006},
protein folding~\cite{Arashiro2,Arashiro3}, propagation of damages in Ising
models~\cite{albano2010}.

The sequence to determine the static exponents from short time dynamics is:
to determine $z$ first, performing
Monte Carlo simulations that mixes initial conditions~\cite{daSilva2002a},
and consider the power law for the cumulant 
\begin{equation}
F_{2}(t)=\frac{\left\langle M^{2}\right\rangle _{m_{0}=0}}{\left\langle
M\right\rangle _{m_{0}=1}^{2}}\sim t^{d/z}\;.  \label{z}
\end{equation}

Once $z$ is calculated, the exponent $\eta =2\beta /\nu $ is calculated
according to $\eta =2\widehat{(\beta /\nu z)}\cdot \widehat{z}$, where $%
\widehat{(\beta /\nu z)}$ was estimated via magnetization decay and $%
\widehat{z}$ from cumulant $F_{2}$.

However, prior to obtaining the critical exponents, we also perform time
dependent MC simulations in order to refine the critical temperatures. These
are based on power laws obtained by finite size scaling analysis of the
magnetization decay from an initially  ordered state (Eq.~\ref{decay_ferro}). This
choice demands a number of runs smaller than other power laws in
non-equilibrium, and so we propose an simple algorithm that spans different
critical values to find the best determination coefficient in linear fit $%
\ln \langle M\rangle $ versus $\ln t$ . This procedure is 
explored in Sec.~\ref{sec:results}, and is  used later to calculate the critical
temperatures for Ising models with   different values of the non-extensivity
parameter $q$ in our new Metropolis prescription.

\section{Generalized master equation}

\label{sec:generalized_master_equation}

In this section, we start recalling the way that the Metropolis algorithm is
obtained from the master equation for spin systems. We point out that the
energy difference by flipping an Ising spin is local,i.e.  it depends only on the
flipped spin. Next, we show a first attempt to generalize the Metropolis
algorithm~\cite{Crokidakis2009,Boer2011}, according to the non-extensive
thermostatistics, introduced by Tsallis~\cite{tsallis_1}. We show that this
generalization does not preserve the spin flip locality. To recover this
locality, we propose a new generalized master equation, which leads to a
different generalization of the Metropolis algorithm.

\subsection{Standard master equation and Metropolis algorithm}

In general, spin systems non-equilibrium dynamics are described by the time
evolution of the probability $P(E,t)$ that, at instant $t$, the system has
an energy $E$. This probability is obtained from the master equation: $%
dP(E^{(a)},t)/dt=\sum\limits_{\sigma _{i}^{(b)}}\{w[\sigma
_{i}^{(b)}\rightarrow \sigma _{i}^{(a)}]P[E^{(b)},t]-w[\sigma
_{i}^{(a)}\rightarrow \sigma _{i}^{(b)}]P[E^{(a)},t]\}$, where $w[\sigma
_{i}^{(b)}\rightarrow \sigma _{i}^{(a)}]$ is the transition rate of the $i-$%
th spin from $\sigma _{i}^{(b)}$ to $\sigma _{i}^{(a)}$. Here, $E^{(b)}$ ($%
E^{(a)}$) is the energy of the system before (after) the transition. As $%
t\rightarrow \infty $, $dP(E,t)/dt=0$ is a necessary condition for equilibrium. 
A sufficient but not necessary condition for equilibrium, known as detailed balance 
condition, supposes a more restricted situation for ocurrence of $dP(E,t)/dt=0$,
i.e., $w[\sigma _{i}^{(b)}\rightarrow \sigma
_{i}^{(a)}]P[E^{(b)}]-w[\sigma _{i}^{(a)}\rightarrow \sigma
_{i}^{(b)}]P[E^{(a)}]=0$, meaning that each term in the summation vanishes.
In this case, $P(E)=P(E,t\rightarrow \infty )$ is the Boltzmann
distribution: $P(E_{j})=e^{-\beta E_{j}}/\sum_{k}e^{-\beta E_{k}}$, where
the summation is over the different energy states and $\beta =(k_{B}T)^{-1}$.

Employing detailed balance requires to find simple prescriptions for
spins systems dynamics, as for example the Metropolis prescription: $%
w[\sigma _{i}^{(b)}\rightarrow \sigma _{i}^{(a)}]=\min \{1,\exp [-\beta
(E^{(a)}-E^{(b)})]\}$. When applied to evolve spin systems, this simple
dynamics reduce to calculate just local energy changes. For instance, the
Ising model in two dimensions has an energy $E^{(b)}=-J\sigma
_{i_{x},i_{y}}^{(b)}S_{i_{x},i_{y}}+\xi $ before the flip of spin $\sigma
_{i_{x},i_{y}}$, located at site indexed by $i_{x}$ and $i_{y}$, where the
local energy change is quantified by 
\[
S_{i_{x},i_{y}}=\sigma _{i_{x}+1,i_{y}}+\sigma _{i_{x}-1,i_{y}}+\sigma
_{i_{x},i_{y}-1}+\sigma _{i_{x},i_{y}+1}
\]%
and the non-local energy is $\xi $, which is obtained excluding the spin $%
\sigma _{i_{x},i_{y}}$, from the calculation. After the spin flip, the
energy is $E^{(a)}=-J\sigma _{i_{x},i_{y}}^{(a)}S_{i_{x},i_{y}}+\xi $ and
the energy change of the system, due to the spin $\sigma _{i_{x},i_{y}}$
flip is simply: 
\begin{equation}
E^{(a)}-E^{(b)}=-J[\sigma _{i_{x},i_{y}}^{(a)}-\sigma
_{i_{x},i_{y}}^{(b)}]S_{i_{x},i_{y}}\;,  \label{eq:energy_difference}
\end{equation}%
which does not depend on the energy of the other spins.

\subsection{Generalized Metropolis algorithm}

The system equilibrium is described by the generalized Boltzmann-Gibbs distribution 
\begin{equation}
P_{1-q}(E_{i})=\frac{[e_{1-q}(-\beta ^{\prime }E_{i})]^{q}}{%
\sum_{i=1}^{\Omega }[e_{1-q}(-\beta ^{\prime }E_{i})]^{q}}\;,  \label{eq:pq}
\end{equation}%
where $\Omega $ is the number of accessible states of the system and $\beta
^{\prime }=\beta /\sum\limits_{i=1}^{\Omega }\{[e_{1-q}(-\beta
E_{i})]^{q}+(1-q)\beta \langle E\rangle _{1-q}\}$, where $\langle E\rangle
_{1-q}=\sum\nolimits_{i=1}^{\Omega }E_{i}P_{1-q}(E_{i})$. Here it is
important to mention that $(k_{B}\beta ^{\prime })^{-1}$ is a scale
temperature that can be used to interpret experimental and computational
experiments. There is a heated ongoing  discussion whether it is  the
physical temperature or not.

The function 
\begin{equation}
e_{\alpha }(x) = \left \{ 
\begin{array}{ll}
(1+\alpha x)^{1/\alpha } &  \mbox{for} \; \alpha x>-1  \\
0                                           & \mbox{otherwise} \; , 
\end{array}
\right.
\end{equation}
is the generalized exponential~\cite{tsallis_2,tiago}. 
For $\alpha \rightarrow 0$, one retrieves the standard
exponential function $e_{0}(x)=e^{x}$. It is this singularity at $\alpha
x>-1 $ that brings up interesting effects such the survival/extinction transitions in one-species population dynamical models~\cite{martinez5}. 
The inverse of the
generalized exponential function is the generalized logarithmic function $%
\ln_{\alpha}(x) = (x^{\alpha} - 1)/\alpha$, which for  $\alpha \rightarrow
0$  leads to the standard logarithm function $\ln_{0}(x) = \ln(x)$.
Notice that the inequality $\alpha x > -1$, for fixed $x$ produces a
limiting value for $\alpha$. This generalized logarithmic function has been
 introduced first  in the context on non-extensive thermostatistics~\cite%
{tsallis_1,tsallis_2} and has a clear geometrical interpretation ss the
area between 1 and $x$ underneath the non-symmetric hyperbole $%
1/t^{1-\alpha} $~\cite{tiago}. It is interesting to notice, that in 1984
Cressie and Read~\cite{cressie} proposed an entropy that would lead to a
generalization of the logarithm function given by : $\ln_{\alpha}(x)/(\alpha + 1)$.
In this case, we would gain the limiting value in $\alpha$ but lose its
geometrical interpretation. 

To recover the additive property of the argument, when multiplying two
generalized exponential functions: $e_{\alpha}(a) e_{\alpha}(b) =
e_{\alpha}(a \oplus_{\alpha} b)$ [$e_{\alpha}(a) / e_{\alpha}(b) =
e_{\alpha}(a \ominus_{\alpha} b)$] and $e_{\alpha}(a) \otimes_{\alpha}
e_{\alpha}(b) = e_{\alpha}(a + b)$ [$e_{\alpha}(a) \oslash_{\alpha}
e_{\alpha}(b) = e_{\alpha}(a - b)$] consider the following algebraic
operators~\cite{nivanen,borges}: 
\begin{eqnarray}
a\oplus _{\alpha }b &=&a+b+\alpha ab  \label{eq_mais} \\
a\ominus _{\alpha }b &=&\frac{a-b}{1+\alpha b}  \label{eq_menos} \\
a\otimes _{\alpha }b &=&\left( a^{\alpha }+b^{\alpha }-1\right) ^{1/\alpha }
\label{eq_vezes} \\
a\oslash _{\alpha }b &=&\left( a^{\alpha }-b^{\alpha }+1\right) ^{1/\alpha }
\; .  \label{eq_dividir}
\end{eqnarray}
Observe that, if $a\ominus _{\alpha }b=0$, then $a=b$ and if $a\otimes
_{\alpha }b=c\otimes _{\alpha }d$, then $a\oslash _{\alpha }c=d\oslash
_{\alpha }b$.

However, in equilibrium, the Ising model prescribes an adapted Metropolis
dynamics that considers a generalized version of exponential function~\cite%
{Crokidakis2009,Boer2011}: 
\begin{equation}
w[\sigma _{i}^{(b)}\rightarrow \sigma _{i}^{(a)}] = \frac{P_{1-q}[E^{(a)}]}{%
P_{1-q}[E^{(b)}]} =\left\{ \frac{e_{1-q}[-\beta^{\prime }E^{(a)}]}{%
e_{1-q}[-\beta^{\prime }E^{(b)}]}\right\}^{q} \; .  \label{Metropolis_I}
\end{equation}

From the generalization of the exponential function in the Boltzmann-Gibbs
weight, the transition rate of Eq.~\ref{Metropolis_I} can be used to
determine the system evolution, as the Metropolis algorithm. Nevertheless,
we stress that in such a choice, the dynamics is not local. Because 
generalized exponential functions are non-additive, a spin flip
introduces a change in the system energy that is spread all over the
lattice. More precisely, consider the Ising model in a square lattice, one
can show that: 
\begin{equation}
\frac{e_{1-q}[-\beta ^{\prime }E^{(a)}]}{e_{1-q}[-\beta ^{\prime }E^{(b)}]}%
=e_{1-q}\{-\beta ^{\prime }[E^{(a)}\ominus _{1-q}E^{(b)}]\}
\label{eq_nonlocal0}
\end{equation}%
or:%
\begin{equation}
\frac{e_{1-q}[-\beta ^{\prime }E^{(a)}]}{e_{1-q}[-\beta ^{\prime }E^{(b)}]}%
\neq e_{1-q}\{-\beta ^{\prime }[E^{(a)}-E^{(b)}]\}\;,  \label{eq_nonlocal}
\end{equation}%
where $E^{(a)}-E^{(b)}$ is given by Eq.~\ref{eq:energy_difference}, which
depends only the spins that directly interact with the flipped spin,
violating the detailed energy balance.

In Refs~\cite{Crokidakis2009,Boer2011}, the authors consider (with no
explanations) the equality in Eq.~\ref{eq_nonlocal}, instead of considering
Eq.~\ref{eq_nonlocal0}. Thus, the detailed energy balance is violated, since
the system is updated following a local calculation of the generalized
Metropolis algorithm of Eq.~\ref{Metropolis_I}.

To correct this problem, one must update the spin system using the
non-locality of Eq.~\ref{Metropolis_I}, which is numerically expensive,
since the energy of the whole lattice must be recalculated due to a simple
spin flip. The other alternative is to require that the transition rate
depends locally in the energy difference of a simple spin flip, which in
turn leads us to a modified master equation. Since the former is very
expensive numerically,  we explore only the latter
alternative which is numerically faster and is able to produce statistically
significant results for fairly large spin systems.

\subsection{Recovering locality in the generalized Metropolis algorithm}

%
%

Based on the operators of Eq.~\ref{eq_mais} to Eq.~\ref{eq_dividir}, we
propose the following generalized master equation: 
\begin{eqnarray}
\frac{dP_{1-q}[E^{(a)}]}{dt}=\sum\limits_{\sigma _{i}^{(b)}} &&w[\sigma
_{i}^{(b)}\rightarrow \sigma _{i}^{(a)}]\otimes _{\tilde{q}%
/q}P_{q}[E^{(b)}]\;\ominus _{\tilde{q}/q}  \nonumber \\
&&w[\sigma _{i}^{(a)}\rightarrow \sigma _{i}^{(b)}]\otimes _{\tilde{q}%
/q}P_{q}[E^{(a)}]\;.  \label{generalized_master_equation}
\end{eqnarray}%
where $P_{q}(E)$ is given by Eq.~\ref{eq:pq}. Here, it is suitable to call $%
\tilde{q}=1-q$ and write the generalized exponentials as a function of $%
\tilde{q}$. In equilibrium, $dP_{1-q}/dt=0$ and a dynamics governed by Eq~%
\ref{eq:pq}.

The detailed balance (a sufficient condition to equilibrium) for the
generalized master equation is 
\begin{widetext}
\begin{equation}
w[\sigma_{i}^{(b)}\rightarrow \sigma_{i}^{(a)}] \oslash _{\tilde{q}/q} w[\sigma_{i}^{(a)}\rightarrow \sigma_{i}^{(b)}] = P_{q}[E^{(a)}] \oslash _{\tilde{q}/q} P_{q}[E^{(b)}] \; ,
\end{equation}
which leads to a new generalized Metropolis algorithm:
\begin{eqnarray}
\nonumber
w(\sigma _{i}^{(b)}\rightarrow \sigma _{i}^{(a)}) & = & \min \left\{ 1,\left[ e_{\tilde{q}}(-\beta ^{\prime }E^{(a)})\right] ^{q}\oslash _{\tilde{q}/q}\left[ e_{\tilde{q}}(-\beta ^{\prime }E^{(b)})\right]^{q}\right\}
                                                                                   =     \min \left\{ 1, \left[ e_{\tilde{q}}(-\beta ^{\prime }(E^{(a)}-E^{(b)}))\right]^{q}\right\} \\ 
                                                                                    & = & \min \left\{ 1,\left[ e_{\tilde{q}}(\beta^{\prime }J\left[ \sigma^{(a)}_{i_{x},i_{y}}-\sigma^{(b)}_{i_{x},i_{y}}\right] S_{i_{x},i_{y}})\right] ^{q}\right\}
\label{Metropolis_II}
\end{eqnarray}
\end{widetext}
and now the transition probability depends only on energy between the read
site and its neighbors, i.e., locality is retrieved.

\section{Generalized Metropolis Algorithm -- Numerical Simulation results}
\label{sec:results}

We have performed Monte Carlo simulations of the square lattice Ising model
in the context of generalized Boltzmann-Gibbs weights. These simulations are
based on two approaches for Metropolis dynamics. The first one (Metropolis
I) is described in Ref.~\cite{Crokidakis2009}, where the nonlocal transition
rate of Eq. (\ref{Metropolis_I}) is used to update the spin system. In the
second approach (Metropolis II), the local transition rate of Eq. \ref%
{Metropolis_II} is used. We separate our results in two different
subsections: the equilibrium simulations and short time critical dynamics.

\subsection{Equilibrium}

In this part we analyze the magnetization $\left\langle m\right\rangle $,
where $\left\langle \cdot \right\rangle $ denotes averages under Monte Carlo
(MC) steps. We perfom MC simulations for $q=0.6$, $q=0.8$ and $q=1.0$. In
the simulations, we have used $L_{\min }=2^{4}=16$ up to $L_{\max }=2^{9}=512
$, with periodic boundary conditions and a random initial configuration of
the spins with $\left\langle m_{0}\right\rangle =0$. Differently
from reported in Ref.~\onlinecite{Crokidakis2009}, where the results have
been obtained after $10^{7}$ MC steps per spin, we have used 
$6.1^{3}$ MC steps per spin, an equilibrium situation consistent
with the one  reported by Newman and Barkema \cite{Newman1999}. This results
in $1.5\cdot 10^{6}-1.5\cdot 10^{9}$ MC steps for the whole
lattice of $16^{2}$  up to $512^{2}$ spins.

\begin{figure}[h]
\begin{center}
\includegraphics[width=\columnwidth]{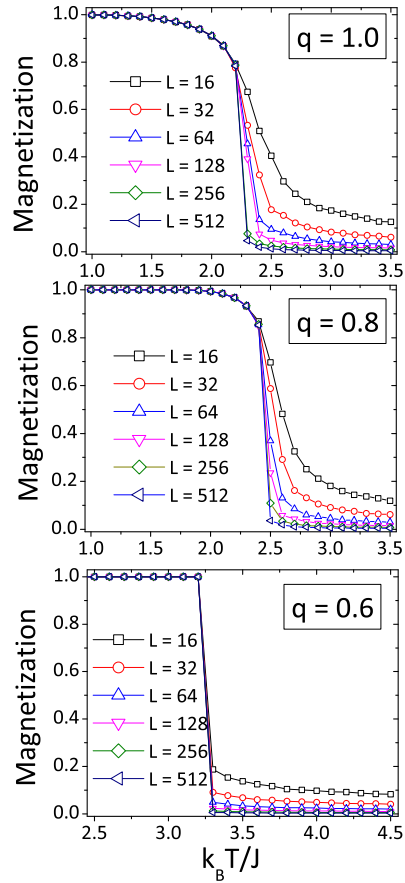}
\end{center}
\caption{System magnetization versus temperature for $q= 1.0$, 0.8 and 0.6.
Using the dynamics based on Metropolis II, we observe phase transitions for
critical values upper to $\log (1+\protect\sqrt{2})/2$ as $q<1$ differently
from previous studies, which are based on Metropolis I.}
\label{fig:magnetization_equilibrio}
\end{figure}

\begin{figure}[h]
\begin{center}
\includegraphics[width=\columnwidth]{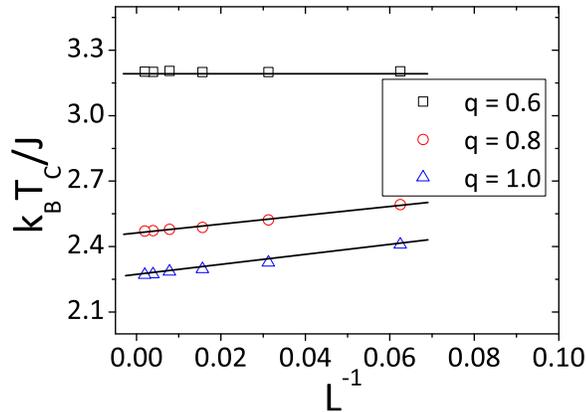}
\end{center}
\caption{Extrapolation ($L\rightarrow \infty $) of critical temperatures for
different $q-$values: $0.6,0.8$ and $1.0$ for the 2d Ising model. }
\label{extrapolation}
\end{figure}

Fig. \ref{fig:magnetization_equilibrio} shows the magnetization curves as
function of critical temperature for different $q$-values. The critical
temperature increases as $q$ decreases. This behavior, using our algorithm
(Metropolis II) differs from the one obtained using the algorithm of Refs.~%
\onlinecite{Crokidakis2009} and~\onlinecite{Boer2011} (Metropolis I). We
stress that both algorithms agree for $q=1$, the usual Boltzmann-Gibbs
weights, converging to the theoretical value $\log (1+\sqrt{2})/2$. In Table %
\ref{table:critical_values_metropolis_I_and_II}, we show the critical
temperature and error obtained from the extrapolation $L\rightarrow \infty $
(see Fig.~\ref{extrapolation}) using both algorithms. These results suggest
a thorough difference among the processes and critical values found between
two the dynamics Metropolis I and II. In Fig.~\ref%
{fig:magnetization_equilibrio}, the curves show phase transitions for
critical values upper to $\log (1+\sqrt{2})/2$ as $q<1$. This differs from
previous studies, which are based on prescription Metropolis I.

Fig.~\ref{fig:magnetization_equilibrio} shows that, differently from the $q=0.8$ and $q=1.0$ cases,
 for $q =0.6$ the discontinuity in the magnetization curve does not depend on system size $L$. 
In fact, in this case, the critical temperature $T_{c}$ does not depend on $L$. 
This effect occurs due to the cutoff of the escort probability distribution
 as reported for Metropolis I \cite{Crokidakis2009} for $q<0.5$.  
For Metropolis II, Fig.~\ref{extrapolation} depicts that $T_{c}$ remains constant for all values of $L^{-1}$, for $q=0.6$. 
For both cases, $q=1.0$ (obviously) and $q=0.8$, we have verified that $\nu \approx 1$ and $\beta \approx 0.125$, 
obtained from the  collapse of the curves $\left\langle M\right\rangle L^{\beta /\nu }$ versus $(T-T_{c})L^{1/v}$. 
This data collapse permits the extrapolation of $k_{B}T_{c}/J$ versus $L^{-1}$, since $\nu \approx 1$ 
for both cases according to Fig.~\ref{extrapolation}. 
In following, we show using non-equilibrium simulations that $2\beta /\nu
\approx 0.25$, for $q=1$ and $q=0.8$, validating the data collapse results
 (see table~\ref{Table:best_values_critical_temperature_MetropolisII}).

\begin{table*}[tbp] \centering%
\begin{tabular}{ccc}
\hline\hline
$q$ & ref. Metropolis I & Metropolis II \\ \hline
0.6 & 1.761(3) & 3.201(1) \\ 
0.8 & 1.891(7) & 2.461(5) \\ 
1.0 & 2.259(11) & 2.262(9) \\ \hline\hline
\end{tabular}%
\caption{Comparison between critical the critical temperature and error, for
the 2d Ising model, obtained from extrapolation $L\rightarrow \infty $ (see
Fig.~\ref{extrapolation}) using the algorithm of Ref~\cite{Crokidakis2009} 
(Metropolis I) and ours (Metropolis II).} \label%
{table:critical_values_metropolis_I_and_II}%
\end{table*}%

Another important question to be formulate is: Can we corroborate the same
behavior in non-equilibrium simulations? Next section, we show results from
MC simulations in non-equilibrium regime under the two dynamics (Metropolis
I and II). We also analyze the critical exponents (dynamic and static) as a
function $q$ from short-time dynamics. We show that short time dynamics
corroborate the behavior predicted by two dynamics suggesting that
Metropolis II indeed presents an increase of critical value as $q$-value
increases different from Metropolis I. Our results suggest that these
technics based on time-dependent simulations can be extended also for $q
\neq 1$, in short range spin models.

\subsection{Short time}

Here we address time dependent MC simulations in the context of so called
short time dynamics. First, to test our methodology, we show that critical
values obtained from non-equilibrium simulations using Metropolis I must
corroborate the critical values obtained in Ref.~\cite{Crokidakis2009},
where MC simulations at equilibrium have been employed. We have checked it.
Nevertheless, as in the equilibrium numerical simulations, we show that
Metropolis II leads to different values from Metropolis I method.

Our algorithm to estimate the critical temperature is divided in two stages.
In the first stage, a coarse grained calculation is performed to estimate
the critical temperature $T_{c}(q)$, for different $q$ values. In the second
stage, one uses the estimated critical temperature obtained in the first
stage to run a non-equilibrium Monte Carlo simulation. We denote the second
state as fine scale stage. In this stage, one determines the dynamical
critical exponent from the short time behavior of the spin system, as
described in Sec.~\ref{sec:critical_dynamics}. Since, even using
non-extensive thermostatistics, the magnetization must behave as a power law 
$\langle M\rangle \sim t^{-\beta /\nu z}$, we conjecture that changing $%
T_{c}(q)$ from $T_{c}^{(\min)}(q)$ up to $T_{c}^{(\max )}(q)$, the best $%
T_{c}(q)$ is the one that leads to the best linear behavior of $\ln \langle
M\rangle $ versus $\ln t$. We have considered $n_{s}=500$ realizations, with
initial magnetization $m_{0}=1$.

From the theoretical critical temperature ($\beta _{c}=J/k_{B}T_{c}=\log (1+%
\sqrt{2})/2$), one allows the temperature to vary in the range from $%
k_{B}T_{c}/J-1$ up to $k_{B}T_{c}/J+1$, setting $k_{B}T/J=[2-\log (1+\sqrt{2}%
)][\log (1+\sqrt{2})]+j\cdot \Delta $, where $\Delta =0.1$ and $j=0,1,\ldots
,20$. This is the coarse grained stage. For each temperature, a linear fit
is performed and one calculates the determination coefficient of fit as: 
\begin{equation}
r=\frac{\sum\limits_{t=1}^{N_{MC}}(\overline{\ln \langle M\rangle }-a-b\ln
t)^{2}}{\sum\limits_{t=1}^{N_{MC}}(\overline{\ln \left\langle M\right\rangle 
}-\ln \langle M\rangle (t))^{2}}
\end{equation}%
and $\overline{\ln \langle M\rangle }=(1/N_{MC})\sum\nolimits_{t=1}^{N_{MC}}%
\ln \langle M\rangle (t)$, where $N_{MC}$ is the number of Monte Carlo
sweeps. In our experiments, we have used $N_{MC}=300$ MC steps. Here, $r=1$
means an exact fit, so that the closer $r$ is from the unity, the better. 
Here, $a$ and $b$ are the linear coefficient and the slope in the
linear fit $\ln \langle M\rangle$ versus $\ln t$, respectively. From $b$, one
estimates the exponent $-\beta \nu /z$. 

In the fine scale stage, we refine the critical temperature $%
k_{B}T_{c}^{(1)}(q)/J$ obtained in the first stage. We use the algorithm
considering $\Delta =0.01$, with $j=0,1, \ldots,20$ considering $%
k_{B}T_{c}^{(2)}(q)/J=k_{B}T_{c}^{(1)}(q)/J-0.1+j\cdot \Delta $, now to find
the best critical temperature in the range $k_{B}T_{c}^{(1)}(q)/J-0.1$ to $%
k_{B}T_{c}^{(1)}(q)/J+0.1$ with precision $\Delta =0.01$.

A natural validation for our algorithm is to reproduce the results obtained
in Ref.~\onlinecite{Crokidakis2009}, in equilibrium, using the Metropolis I
approach, for a specific $q$ value, considering our MC non-equilibrium
simulations. For instance, for $q=0.70$, one has at equilibrium $%
k_{B}T_{c}/J=1.891(7)$, in Ref.~\cite{Crokidakis2009}. After two stages, our
algorithm produces $k_{B}T_{c}^{(2)}/J=1.889$, validating our numerical code.

Next, we use the algorithm with the following values: $q=0.70$, $0.75$, $0.80
$, $0.85$, $0.90$, $0.95$ and $1.00$, in the equilibrium situation. In Table~%
\ref{table:coarse_grained_incorrect}, we show our results for the first
stage (coarse grained) using Metropolis I prescription. The values of the
determination coefficient $\alpha $ of the linear fit $\ln \langle M\rangle $
versus $\ln t$ are presented for different $q$ values. The highest values
(in bold) correspond to best critical temperature found in the first stage.
For example: for $q=0.75$, we have that best $\alpha $ value is 0.940558872,
which corresponds to $k_{B}T_{c}^{(1)}(q)/J=1.86918531$. 
 
In Table~\ref{table:coarse_grained_incorrect}, the symbol  ``--'' corresponds to situations
 where the computation of slopes is not possible, due to large
 deviations in magnetization.

\begin{table*}[tbp] \centering%
\begin{tabular}{c|ccccccc}
\hline\hline
$k_{B}T_{c}^{(1)}/J$ & $q=0.70$ & $q=0.75$ & $q=0.80$ & $q=0.85$ & $q=0.90$
& $q=0.95$ & $q=1.00$ \\ \hline\hline
$T^{\ast }-0.6$ & 0.644974839 & 0.544533372 & 0.433158672 & 0.376097284 & 
0.326410245 & 0.282451042 & 0.263459517 \\ 
$T^{\ast }-0.5$ & \textbf{0.998967222} & 0.872350719 & 0.65553425 & 
0.492266999 & 0.392016875 & 0.336915743 & 0.306414971 \\ 
$T^{\ast }-0.4$ & 0.858060019 & \textbf{0.940558872} & \textbf{0.979041762}
& 0.731374836 & 0.500535816 & 0.396550709 & 0.341806493 \\ 
$T^{\ast }-0.3$ & 0.822853648 & 0.82063843 & 0.90326648 & \textbf{0.999101612%
} & 0.773207193 & 0.535343676 & 0.409044416 \\ 
$T^{\ast }-0.2$ & -- & -- & 0.833548876 & 0.885834994 & \textbf{0.998950355}
& 0.788883669 & 0.547803953 \\ 
$T^{\ast }-0.1$ & -- & -- & -- & 0.836458324 & 0.882565862 & \textbf{%
0.999817616} & 0.776353951 \\ 
$T^{\ast }=\log (1+\sqrt{2})/2$ & -- & -- & -- & -- & 0.817075651 & 
0.897612219 & \textbf{0.997114577} \\ 
$T^{\ast }+0.1$ & -- & -- & -- & -- & -- & 0.82435859 & 0.916225167 \\ 
\hline\hline
\end{tabular}%
\caption{Coarse grained Stage for Metropolis I. 
The values of determination coefficient $\alpha$ of the linear fit  $\ln \langle M\rangle$ versus $\ln t$ for different $q$ values. 
The highest values are in bold and correspond to best critical temperature found at first stage (coarse grained). 
For example: for $q$=0.75,  the best $r$ is 0.940558872, which corresponds to $k_{B}T_{c}^{(1)}/J$=1.86918531.}%
\label{table:coarse_grained_incorrect}%
\end{table*}%

After the refinement (second stage), the best values found for the critical
temperatures using Metropolis I prescription, for different $q-$values are
presented in the first line of Table~\ref%
{Table:best_values_critical_temperature_MetropolisI}. In Fig.~\ref%
{fig:decay_magnetization}, for $q = 0.70$ and 0.85, we show the
magnetization decays as the power law: $M(t)\sim t^{-\beta /\nu z}$, for the
critical temperature estimated using: our (Metropolis II) and Metropolis I
algorithms. Also, we show the plots considering MC simulations for $%
T_{c}+\delta $ and $T_{c}-\delta $, with $\delta =0.05$.

\begin{figure}[h]
\begin{center}
\includegraphics[width=\columnwidth]{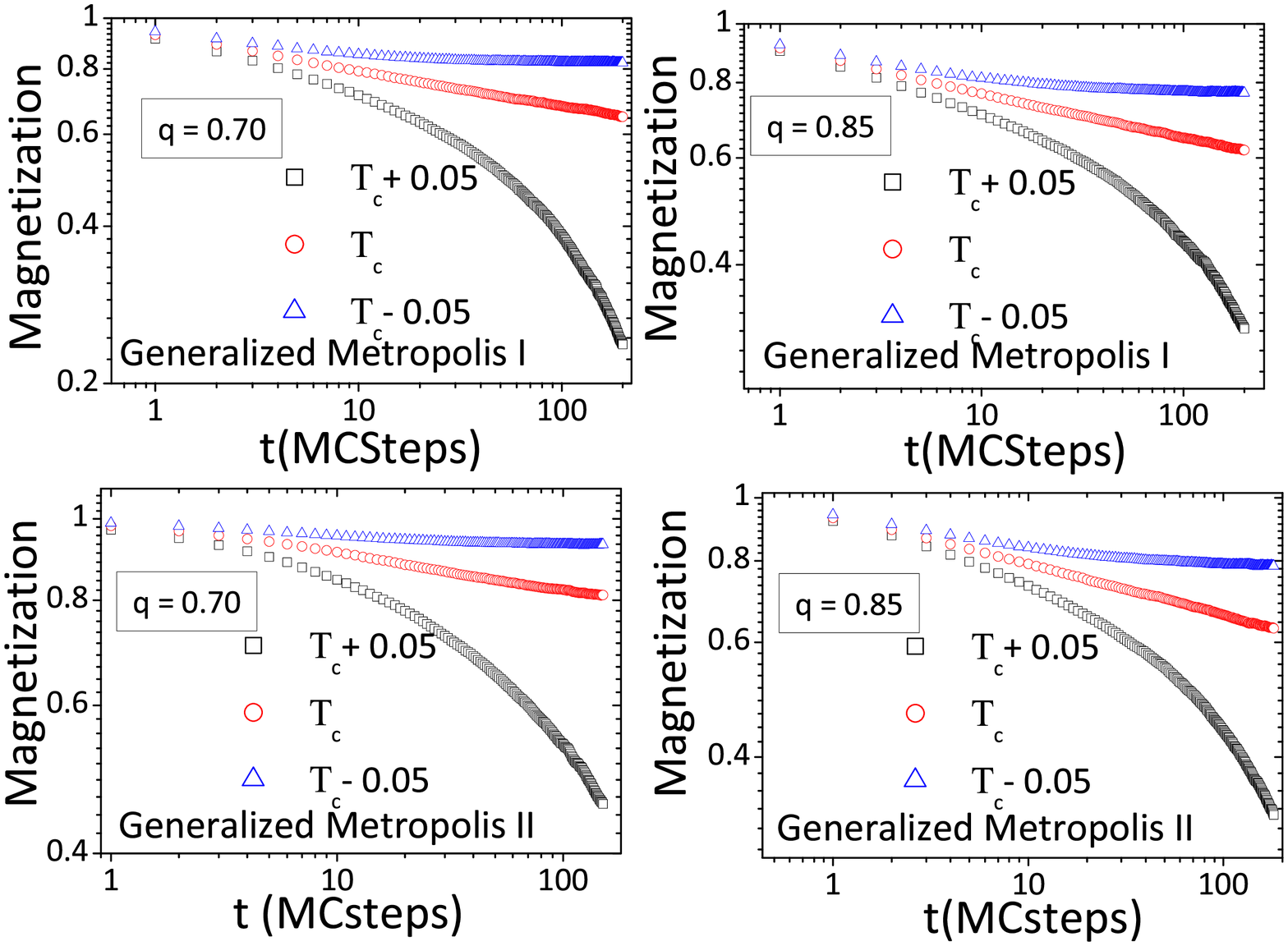}
\end{center}
\caption{Decay of magnetization according to the power law: $M(t)\sim t^{-%
\protect\beta /\protect\nu z}$ in the critical temperature found by the
considered algorithms (circle red points), for $q=0.70$ and 0.85. We also
show the plots considering MC simulations for $T_{c}+\protect\delta $ and $%
T_{c}-\protect\delta $. It was used $\protect\delta =0.05$. The upper
(lower) plots correspond to Metropolis I (II) algorithm.}
\label{fig:decay_magnetization}
\end{figure}

We use the same procedure to find the critical temperatures for prescription
Metropolis II. We find very different results, when compared with that ones
obtained with Metropolis I. Similarly to Table \ref%
{table:coarse_grained_incorrect}, we show the results using the Metropolis
II prescription in Table~\ref{table:corse_grained_correct}. The values are
smaller than the ones found with Metropolis I prescription. However, they
match as $q\rightarrow 1$, which validates the numerical procedure.

\begin{table*}[tbp] \centering%
\begin{tabular}{l|lllllll}
\hline\hline
$q$ & 0.70 & 0.75 & 0.80 & 0.85 & 0.90 & 0.95 & 1.00 \\ \hline\hline
$k_{B}T_{c}^{(2)}/J$ & 1.77(1) & 1.82(1) & 1.89(1) & 1.97(1) & 2.07(1) & 
2.17(1) & 2.27(1) \\ 
$\beta /\nu z$ & 0.060(4) & 0.062(7) & 0.078(5) & 0.082(2) & 0.100(5) & 
0.094(4) & 0.057(3) \\ 
$z$ & 2.13(4) & 2.15(5) & 2.12(4) & 2.09(3) & 2.10(3) & 2.11(6) & 2.15(3) \\ 
$\theta $ & 0.18(4) & 0.14(4) & 0.22(7) & 0.17(3) & 0.04(6) & 0.17(3) & 
0.19(4) \\ 
$\eta $ & 0.25(2) & 0.27(3) & 0.33(2) & 0.34(1) & 0.42(2) & 0.40(2) & 0.25(1)
\\ 
$r$ & 0.998568758 & 0.998915473 & 0.999342437 & 0.999458152 & 0.999589675 & 
0.999718708 & 0.999206853 \\ \hline\hline
\end{tabular}%
\caption{Critical temperature and exponents obtained for different $q$ values for prescription Metropolis I. 
The exponents where obtained performing simulations for the estimated critical temperatures and were based 
on power laws previously described in short time regime. The last line we show the $r$ value for the 
best fits in the second stage (fine scale)} \label%
{Table:best_values_critical_temperature_MetropolisI}%
\end{table*}%

\begin{table*}[tbp] \centering%
\begin{tabular}{c|ccccccc}
\hline\hline
$k_{B}T_{c}^{(1)}/J$ & $q=0.70$ & $q=0.75$ & $q=0.80$ & $q=0.85$ & $q=0.90$
& $q=0.95$ & $q=1.00$ \\ \hline\hline
$T^{\ast }-0.1$ & -- & 0.369354816 & 0.393077206 & 0.473151224 & 0.555233203
& 0.664145691 & 0.754299904 \\ 
$T^{\ast }=\log (1+\sqrt{2})/2$ & -- & 0.439789081 & 0.50595381 & 0.658725599
& 0.829581264 & \textbf{0.952694449} & \textbf{0.997206853} \\ 
$T^{\ast }+0.1$ & -- & 0.579484235 & 0.756653731 & \textbf{0.959530136} & 
\textbf{0.995694839} & 0.951627799 & 0.91284158 \\ 
$T^{\ast }+0.2$ & 0.601895662 & 0.836896384 & \textbf{0.999315534} & 
0.932708995 & 0.869271327 & 0.848547425 & 0.838519074 \\ 
$T^{\ast }+0.3$ & 0.844382176 & \textbf{0.989767198} & 0.875131078 & 
0.833730495 & 0.831169795 & 0.799709762 & -- \\ 
$T^{\ast }+0.4$ & \textbf{0.989716381} & 0.847004746 & 0.812106454 & -- & --
& -- & -- \\ 
$T^{\ast }+0.5$ & 0.828738110 & 0.787203767 & -- & -- & -- & -- & -- \\ 
$T^{\ast }+0.6$ & 0.842827863 & -- & -- & -- & -- & -- & -- \\ \hline\hline
\end{tabular}%
\caption{Coarse grained Stage for Metropolis II - The values of determination coeficient $r$ of the linear fit  $\ln \langle M \rangle$ versus $\ln t$, for different $q$ values. 
As in Table~\ref{table:coarse_grained_incorrect}, the highest values are in bold correspond to best critical temperature found at first stage (coarse grained).}
\label{table:corse_grained_correct}%
\end{table*}%

Similarly, the best results after the fine scale refinement (second stage)
are shown in the first line of Table \ref%
{Table:best_values_critical_temperature_MetropolisII}.

\begin{table*}[tbp] \centering%
\begin{tabular}{l|lllllll}
\hline\hline
$q$ & 0.70 & 0.75 & 0.80 & 0.85 & 0.90 & 0.95 & 1.00 \\ \hline\hline
$k_{B}T_{c}^{(2)}/J$ & 2.66(1) & 2.55(1) & 2.47(1) & 2.41(1) & 2.36(1) & 
2.31(1) & 2.27(1) \\ 
$\beta /\nu z$ & 0.019(5) & 0.039(5) & 0.060(4) & 0.094(6) & 0.116(7) & 
0.075(4) & 0.057(3) \\ 
$z$ & 1.97(4) & 2.02(3) & 2.10(3) & 2.09(3) & 2.09(6) & 2.20(4) & 2.15(3) \\ 
$\theta $ & 0.43(3) & 0.21(7) & 0.22(3) & 0.11(4) & 0.16(5) & 0.13(3) & 
0.19(4) \\ 
$\eta $ & 0.07(2) & 0.16(2) & 0.25(2) & 0.39(3) & 0.48(3) & 0.33(2) & 0.25(1)
\\ 
$r$ & 0.994455464 & 0.998375667 & 0.999227272 & 0.999226704 & 0.99928958 & 
0.99925661 & 0.997206853 \\ \hline\hline
\end{tabular}%
\caption{Critical temperature and exponents obtained for different $q$ values for the Metropolis II algorithm. 
The exponents were obtained performing simulations for the estimated critical temperatures. 
They were based  on power laws described in Sec.~\ref{sec:critical_dynamics}. 
The last line we show the $r$ value for the  best fits in the second stage (fine scale)}
\label{Table:best_values_critical_temperature_MetropolisII}%
\end{table*}%

The magnetization decay obtained by the Metropolis II algorithm depicted in
Fig.~\ref{fig:decay_magnetization}. After obtaining these estimates for the
critical temperatures, we perform short time simulations to obtain the
critical dynamic exponents $z$ and $\theta $ and the static one $\eta
=2\beta /\nu $, using the power laws of Sec.~\ref{sec:critical_dynamics}.
Here, we calculated $\theta $ from time correlation $C(t)=\langle
M(t)M(0)\rangle $. Tom\'{e} and de~Oliveira~\cite{Tania1998} showed that
correlation behaves as $C(t)\sim t^{\theta }$, where $\theta $ is exactly the
same exponent from initial slope of magnetization from lattices prepared
with initial fixed magnetization $m_{0}$. The advantage of this method is
that we repeat $N_{s}$ runs, but the lattice does not require a fixed
initial magnetization. It is enough to choose the spin with probability 1/2.
i.e., $m_{0}=0$ in average. This method does not require the extrapolation $%
m_{0}\rightarrow 0$.

\begin{figure}[h]
\begin{center}
\includegraphics[width=\columnwidth]{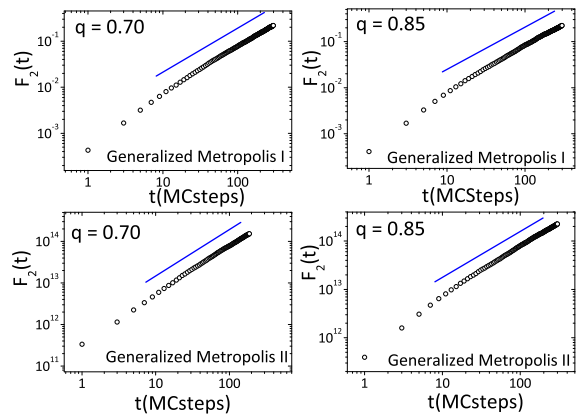}
\end{center}
\caption{Dynamic cumulant $F_{2}(t)$ versus $t$ in log scale. The slope
gives $d/z$ which supplies the $z$-value. Both prescriptions (Metropolis I e
II) are studied. }
\label{f2}
\end{figure}

\begin{figure}[h]
\begin{center}
\includegraphics[width=\columnwidth]{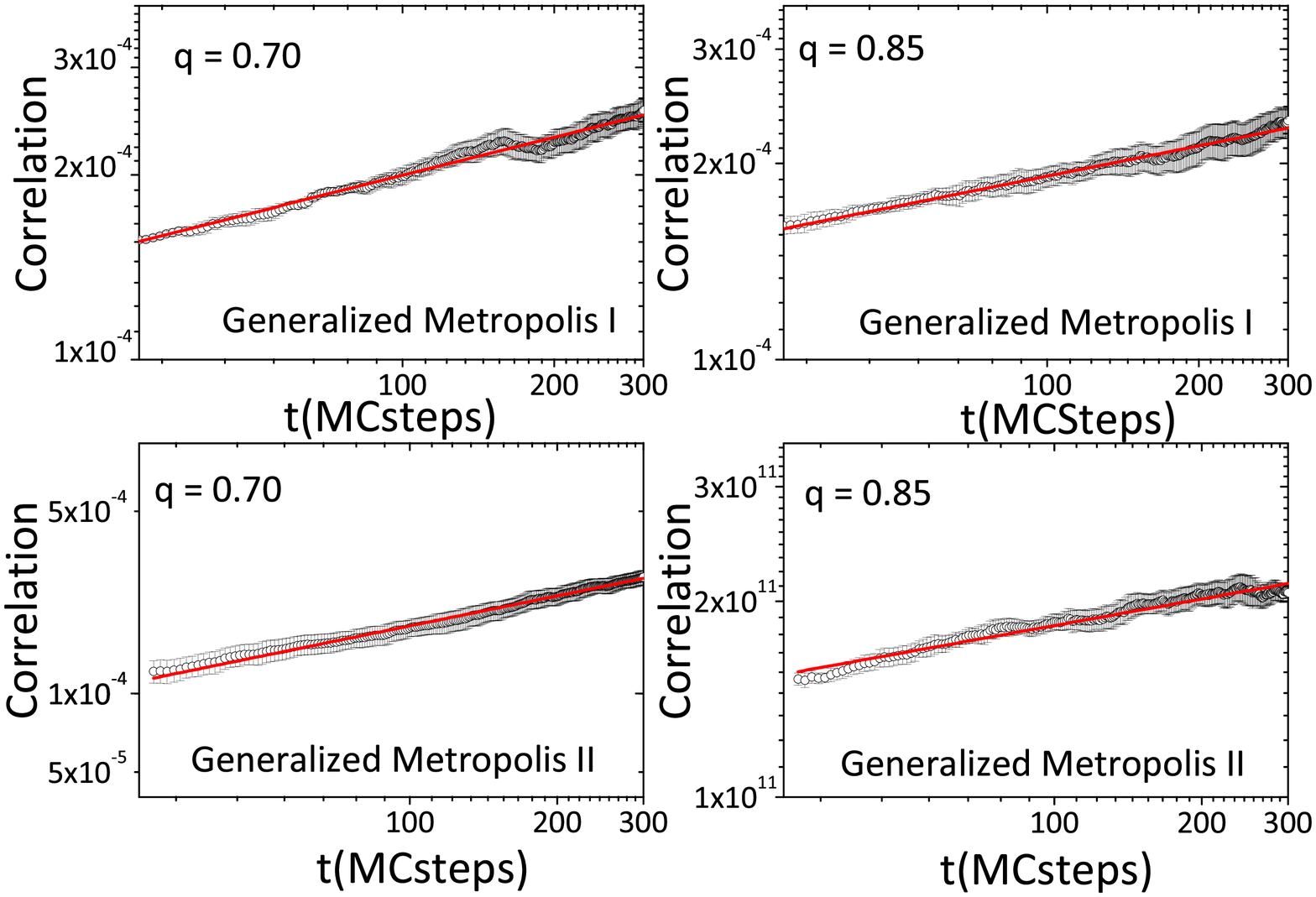}
\end{center}
\caption{Time correlation of magnetization $C(t)=\left\langle
M(t)M(0)\right\rangle $ for two prescriptions: Metropolis I and II.}
\label{correlation}
\end{figure}

Figs.~\ref{f2} and~\ref{correlation} depict plots of time evolving of $F_{2}$
of Eq.~\ref{z} and $C(t)$ as a function of $t$ for the different Metropolis
algorithm.

To obtain the exponents consider the following steps. Firstly, in
simulations that start from the ordered state $m_{0}=1$ and $L=512$,
calculate the slope $\beta/\nu z$ of the linear fit of $\ln \langle M(t)
\rangle $ as a function of $\ln t$. The error bars are obtained, via running
simulations for $N_{bin}=5$, calculating $\langle M(t) \rangle $ that each
for seed, with $N_{run}=400$ runs.

Once we have calculated $\beta /\nu z$, we estimate $z$ taking the slope in
log-log plot $\ln F_{2}$ versus $\ln t$. We used $N_{s}=3000$ different runs
starting from random spins configurations with $m_{0}=0$ for time series $%
\langle M(t)^{2}\rangle $ $\times \ t$ and the same number of runs for time
series $\langle M(t)\rangle $ $\times \ t$ starting from $m_{0}=1$(ordered
state). Similarly, we repeated the numerical experiment for $N_{bin}=5$
different seeds to obtain the uncertainties. In two dimensional systems, the
slope is $\phi =2/z$ (see Eq.~\ref{z}) and so $z$ is calculated according to 
$\widehat{z}=2/\widehat{\phi }$ and the uncertainty in $z$ is obtained by
relation$\ \sigma _{z}=(2/\phi ^{2})\sigma _{\phi }$. Here the $\widehat{%
\bullet }$ denotes the amount estimated from $N_{bin}=5$ different seeds.
Once $z$ is calculated, the exponent $\eta =2\beta /\nu $ is calculated
according to $\eta =2\widehat{(\beta /\nu z)}\cdot \widehat{z}$, where $%
\widehat{(\beta /\nu z)}$ was estimated via magnetization decay and $%
\widehat{z}$ from cumulant $F_{2}$. The exponent $\theta $ was similarly
obtained performing $N_{s}=3000$ different runs to evolve the time series of
correlation $C(t)$ and estimating directly the slope in this case.

Tables~\ref{Table:best_values_critical_temperature_MetropolisI} and~\ref%
{Table:best_values_critical_temperature_MetropolisII} show results for the
critical exponents obtained with the two algorithms. We do not observe a
monotonic behavior of the critical exponents as function of $q$ in either
case but on the other hand for both cases we cannot assert, for example,
that $z\in \lbrack 2.09,2.15]$ (Metropolis I) and $z\in \lbrack 1.97,2.20]\ $%
(Metropolis II) or even other exponents do not change for $q<1$ which
implies that we cannot simply extrapolate the critical properties from $q=1$
to $q<1$.

\section{Conclusions}

\label{sec:conclusions}

In the non-extensive thermostatistics context, we have proposed a
generalized master equation leading to a generalized Metropolis
algorithm. This algorithm is local and satisfies the detailed energy balance
to calculate the time evolution of spins systems. We calculate the critical
temperatures using the generalized Metropolis dynamics, via equilibrium and
non-equilibrium Monte Carlo simulations.

We have obtained the critical parameters performing Monte Carlo simulations
in two different ways. Firstly, we show the phase transitions from curves $%
\left\langle M\right\rangle $ versus $k_{B}T/J$, considering the
magnetization averaging, in equilibrium, under different MC steps. Next, we
use the short time dynamics, via relaxation of magnetization from samples
initially prepared of ordered or disordered states, i.e., time series of
magnetization and their moments averaged over initial conditions and over
different runs.

We have also studied the Metropolis algorithm of Refs.~\cite%
{Crokidakis2009,Boer2011}. We show that  it does not preserve locality neither
the detailed energy balance in equilibrium. While our non-equilibrium simulations
corroborate  results of Refs.\cite{Crokidakis2009,Boer2011} when we use 
their extension of the Metropolis algorithm (Metropolis I),  the
exponents and critical temperatures obtained are very different when we use
our prescription (Metropolis II). When the extensive case is considered,
both methods lead to the same expected values.

Simultaneously, we have developed a methodology to refine the determination
of the best critical temperature. This procedure is based on optimization of
the power laws of the magnetization function that relaxes from ordered state
in log scale, via of maximization of determination coefficient of the linear
fits. This approach can be extended for other spin systems, since their
general usefulness.

For a more complete elucidation about existence of phase transitions
for $q\neq 1$, we have performed simulations for small systems MC
simulations, recalculating the whole lattice energy in each simple spin
flip, according to Metropolis I algorithm only to check the variations on
the critical behavior of the model. Notice that this does not apply to
Metropolis II algorithm, since it has been designed to work as the standard
Metropolis one. Our numerical results show discontinuities in the
magnetization, but no finite size scaling, corroborating the results of 
Ref.~\cite{Penna1999}, which used the broad histogram technics to show that no
phase transition occurs for $q\neq 1$ using Metropolis I
algorithm.

It is important to mention that only Metropolis I \cite{Crokidakis2009,Boer2011} 
shows inconsistence on critical phenomena of model since global and 
local simulation schemes leads to different critical properties. Metropolis II 
overcomes this problem since local and global prescriptions are the same even for 
$q\ne 1$”. Broad histogram method works with a non-biased random walk that explore 
the configuration space, leading to a phase transition suppression for $q\ne 1$ \cite{Penna1999}. 
Nevertheless this algorithm must also be adapted to deal with the generalized Boltzmann 
weight in the same way the master equation needed to be modified. This is out of the scope 
of the present paper but this issue will be treated in a near future. 


\section*{Acknowledgements}


The authors are partly supported by the Brazilian Research Council CNPq
under grants 308750/2009-8, 476683/2011-4, 305738/2010-0, and 476722/2010-1.
Authors also thanks Prof. U. Hansmann for carefully reading this manuscript,
as well as CESUP (Super Computer Center of Federal University of Rio Grande
do Sul) and Prof. Leonardo G. Brunet (IF-UFRGS) for the available
computational resources and support of Clustered Computing
(ada.if.ufrgs.br). Finally we would to thank the excellent quality of reviews
of the anonymous referees.

\end{document}